\documentclass[showpacs,twocolumn,superscriptaddress]{revtex4}
\usepackage{amsmath,amssymb,mathrsfs,graphicx}

\allowdisplaybreaks[4]

\begin{document}
\title{Spin and orbital fluctuations in non-equilibrium transport through\\
  quantum dots: A renormalisation-group analysis}  
\author{S. Y. M\"uller}
\affiliation{Institut f\"ur Theorie der statistischen Physik and 
  JARA-Fundamentals of Future Information Technology, RWTH Aachen, 
  52056 Aachen, Germany}
\author{V. Koerting}
\affiliation{Nano-Science Center, Niels Bohr Institute, Universitetsparken 5, 
  2100 Copenhagen, Denmark}
\affiliation{Niels Bohr International Academy, Niels Bohr Institute,
  Blegdamsvej 17, 2100 Copenhagen, Denmark}
\author{D. Schuricht}
\affiliation{Institut f\"ur Theorie der statistischen Physik and 
  JARA-Fundamentals of Future Information Technology, RWTH Aachen, 
  52056 Aachen, Germany}
\author{S. Andergassen}
\affiliation{Institut f\"ur Theorie der statistischen Physik and 
  JARA-Fundamentals of Future Information Technology, RWTH Aachen, 
  52056 Aachen, Germany}
\date{\today}

\pagestyle{plain}
\begin{abstract}
  We study non-equilibrium current and occupation probabilities of a
  two-orbital quantum dot. The couplings to the leads are allowed to
  be asymmetric and orbital dependent as it is generically the case in
  transport experiments on molecules and nanowires.  Starting from a
  two-orbital Anderson model, we perform a generalised
  Schrieffer-Wolff transformation to derive an effective Kondo model.
  This generates an orbital potential scattering contribution which is
  of the same order as the spin exchange interaction.  In a first
  perturbative analysis we identify a regime of negative differential
  conductance and a cascade resonance in the presence of an external
  magnetic field, which both originate from the non-equilibrium
  occupation of the orbitals.  We then study the logarithmic
  enhancement of these signatures by means of a renormalisation-group
  treatment. We find that the orbital potential scattering
  qualitatively changes the renormalisation of the spin exchange
  couplings and strongly affects the differential conductance for
  asymmetric couplings.
\end{abstract}
\maketitle


\section{Introduction}

Over the past decade the fast progress in nanofabrication allowed the
observation of single electron tunneling as well as Coulomb blockade,
cotunneling and Kondo signatures in quantum dot devices. Initially
these dots were gate-controlled isolated islands in two-dimensional
electron gases like GaAs~\cite{Hanson}. Recently a new trend has
evolved towards studying molecules in gated junctions. Starting from
small molecules~\cite{ndcexp,Osorio} and bucky balls~\cite{Roch} also carbon
nanotubes~\cite{Sapmaz,jens} and InAs nanowires were
studied~\cite{Jespersen}.  In contrast to gate-defined quantum dots
these molecules and nanowires possess a rich spectrum of internal
degrees of freedom, which significantly influence non-equilibrium
transport through these systems. In particular, the overlaps of the
orbitals of a molecule with the states in the leads vary
significantly, which results in asymmetric and orbital-dependent
tunneling amplitudes.

Parallel to the experimental progress there have been numerous theoretical
studies on non-equilibrium transport through nanostructures~\cite{review}. One
of the main challenges consists in the new energy scale introduced by the bias
voltage, which is in general of the same order as the internal energy scales
of the system.  The understanding of non-equilibrium phenomena in the presence
of strong correlations represents an outstanding task in condensed matter
physics which requires the development of powerful theoretical tools.  Recent
advances include various renormalisation group (RG)
methods~\cite{Rosch,RGmethods,S} introduced for the study of quantum dots
mainly dominated by spin fluctuations.

\begin{figure}[b]
\centering
\includegraphics[width=0.75\linewidth,clip]{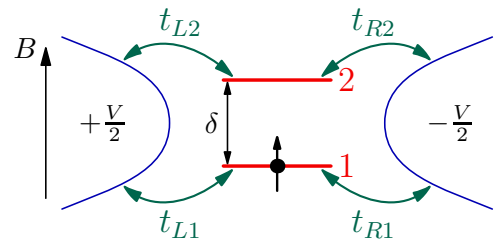}
\caption{(Colour online) Sketch of the two-orbital Anderson model. The
  energies of the two orbitals are split by $\delta$. The orbitals are
  coupled to free fermionic leads held at different chemical
  potentials $\pm V/2$. The system is assumed to be in the regime of
  single occupancy and the dot orbitals are subject to an external
  magnetic field $B$.}
\label{fig:fig1}
\end{figure}

Motivated by the experimental 
developments~\cite{ndcexp,Osorio, Roch, Sapmaz, jens,Jespersen} 
we will study a quantum dot
possessing multiple orbitals and asymmetric tunneling amplitudes. A
previous perturbative study~\cite{Schmaus} showed that such systems
generically exhibit interesting non-equilibrium features like the
so-called cascade effect and the appearance of a negative differential
conductance (NDC). Our main focus is the analysis of the logarithmic
enhancements of these features due to the Kondo effect and the
interplay of spin and orbital degrees of freedom. Even without
spin-orbit interaction the renormalisation of spin and orbital
scattering processes can influence each other in multi-orbital quantum
dots.  The minimal model to observe these features is provided by a
two-orbital quantum dot with finite orbital splitting, an external
magnetic field, and differing tunneling amplitudes (see
fig.~\ref{fig:fig1}).

We consider the Coulomb blockade region with a single electron on the
dot. In this situation the orbital or spin degree of freedom can be
changed by elastic or inelastic cotunneling processes.  Using a
Schrieffer-Wolff (SW) transformation~\cite{SW} we derive a Kondo spin
exchange as well as an orbital potential scattering (PS) interaction,
which have the same strength and therefore must be treated on an equal
footing. The orbital PS term includes two processes, one involving a
change of the orbital index and one describing transport via the empty
orbital.

We first study the orbital occupations and the differential
conductance to second order in perturbation theory (PT), which
provides a reliable and comprehensive picture of the considered
non-equilibrium physics.  In the main part of the paper we perform an
RG analysis with frequency-dependent couplings~\cite{Rosch} and study
the effect of the orbital PS.  This method was recently used to study
cotunneling transport through carbon nanotubes modelled by a
two-orbital Anderson dot with two electrons~\cite{jens}. While for
this situation orbital PS contributions are expected to play an
important role, the impact of an empty orbital remains to be assessed.
Furthermore, transport through a double quantum dot coupled to four
different leads was investigated in ref.~\cite{held}. In this work NDC
with respect to one pair of the leads was observed and a detailed RG
analysis of the transport properties as a function of the orbital
splitting was performed.

In our set-up (see fig.~\ref{fig:fig1}) we study the dependence of the cascade
resonance and the NDC on both the coupling asymmetry and the external
magnetic field.  By including the logarithmic corrections to the inelastic
cotunneling processes we further show that the interplay between spin and
orbital scattering is most pronounced for strong asymmetric couplings to the
reservoirs. The detected enhancement of the NDC reveals the importance
of Kondo correlations for the appearance of this effect, whose signatures can
be probed in transport experiments~\cite{ndcexp}.

\section{Model and SW transformation}

We start from a two-orbital Anderson model sketched in
fig.~\ref{fig:fig1}.  The Hamiltonian is given by $H=H_{\rm
  leads}+H_{\rm dot}+H_{\rm int}$. In second quantisation the
fermionic leads are described by $H_{\rm leads}=\sum_{\alpha
  k\sigma}(\varepsilon_{\alpha k\sigma}-\mu_\alpha)c_{\alpha
  k\sigma}^{\dagger} c_{\alpha k\sigma}$ with $\alpha=L,R$, the
chemical potentials $\mu_{\alpha}=\pm eV/2$ and the density of states
$N(\omega)=N_0\Theta(D_0-|\omega|)$, $N_0=1/(2D_0)$.  The dot
Hamiltonian reads
\begin{equation}
H_{\rm dot}=\sum_{as}\varepsilon_{as}d_{as}^{\dagger} d_{as}+
\frac{U}{2}
\sum_{as}d_{as}^{\dagger} d_{as}\Biggl(\sum_{bs'}d_{bs'}^{\dagger} 
d_{bs'}-1\Biggr)
\label{eq:AM2}
\end{equation}
with the orbital index $a=1,2$, the spin index $s=\pm1$, and an
orbital-independent local Coulomb interaction.  The energies of the
orbitals are $\varepsilon_{1s}=\varepsilon_1-\frac{1}{2}s g\mu_BB$ and
$\varepsilon_{2s}=\varepsilon_{1s}+\delta$ with an external magnetic
field $B$. We stress that the orbital splitting $\delta$ is assumed to
be finite. The orbitals are coupled to the leads by $H_{\rm
  int}=\sum_{\alpha a\sigma}(t_{\alpha a}c_{\alpha\sigma}^{\dagger}
d_{a\sigma}+\text{h.c.})$ with spin-independent hopping amplitudes
$t_{\alpha a}$ and $c_{\alpha\sigma}^{\dagger}=\sum_kc_{\alpha
  k\sigma}^{\dagger}$.

In the following we will consider the regime with exactly one electron
on the dot, i.e. $\sum_{as}n_{as}=1$ with $n_{as}=d_{as}^{\dagger}
d_{as}$.  We use a generalised SW transformation to derive an
effective Kondo model $H_{\rm K}+H_{\rm PS}$ in the limit $\delta\ll
U$. We find the conventional Kondo spin exchange interaction in its
two-orbital form
\begin{equation}
H_{\rm K}=\sum_{\alpha\beta}\sum_{ab}J^{\beta\alpha}_{ba}
\vec{S}_{ba}\cdot\vec{s}_{\beta\alpha},
\label{eq:SW29}
\end{equation} 
with $J^{\beta\alpha}_{ba}=2t_{\beta a}t_{\alpha
  b}^*(1/(\varepsilon_b+U)-1/\varepsilon_a)$, $\vec{S}_{ba} =
\frac{1}{2}\sum_{ss'} d^\dag_{bs} \vec{\tau}_{ss'} d_{as'}$, and
$\vec{s}_{\beta\alpha} = \frac{1}{2}\sum_{\sigma\sigma'} c^\dag_{\beta\sigma}
\vec{\tau}_{\sigma\sigma'} c_{\alpha\sigma'}$.  In an experimental
set-up the overlaps between the states on the dot and in the leads
will strongly depend on the lead and orbital indices, thus leading to a
strong asymmetry as well as orbital dependence of the hopping amplitudes
$t_{\alpha a}$ and the Kondo couplings $J^{\beta\alpha}_{ba}$.  In
addition to \eqref{eq:SW29} the SW transformation generates a
spin-independent \textit{orbital} PS term
\begin{eqnarray}
H_{\rm PS}=\frac{1}{4}\sum_{\alpha\beta\sigma}
c_{\beta\sigma}^{\dagger} c_{\alpha\sigma}
\sum_{as}\left(J^{\beta\alpha}_{\bar{a}a}
d_{\bar{a}s}^{\dagger} d_{as}-J^{\beta\alpha}_{\bar{a}\,\bar{a}}
d_{as}^{\dagger} d_{as}\right),
\label{eq:SW30}
\end{eqnarray}
where $\bar{a} = 2 (1)$ if $a = 1 (2)$.  The second term $\propto
J_{\bar{a}\bar{a}}^{\beta\alpha}$ describes cotunneling transport
across the empty orbital, e.g. orbital $2$ if orbital $1$ is occupied. The
first term $\propto J^{\beta\alpha}_{\bar{a}a}$ is a spin-conserving
process which changes the orbital state of the dot; we therefore
expect a significant influence on the current at voltages
$V\sim\delta$.  We emphasise that \eqref{eq:SW30} is of the same order as
the Kondo interaction \eqref{eq:SW29} and hence cannot be
neglected. In particular, we show below that the orbital PS term
qualitatively changes the RG flow of the Kondo couplings.  In order to
distinguish the Kondo and orbital PS contributions we will denote the
couplings of the latter by $P^0$ ($\equiv
J_{\bar{a}\bar{a}}^{\beta\alpha}$) and $P^x$ ($\equiv
J^{\beta\alpha}_{\bar{a}a}$) in the following.  Furthermore, we study
the system at the particle-hole symmetric point, i.e. $\varepsilon_1 =
- U/2 - \delta/2$.  In this case the conventional single-orbital PS
contribution ($\propto 1/(\varepsilon_a+U) + 1/\varepsilon_a$) can be
neglected ($\delta\ll U$ has already been assumed in the SW
transformation).  

To simplify the notations we introduce a generalised vertex
$\mathcal{V}$ via
\begin{equation}
H_{\rm K}+H_{\rm PS}=\sum_{\alpha\beta ab }\sum_{ss'\sigma\sigma'}
\mathcal{V}_{bs';as}^{\beta\sigma';\alpha\sigma}d_{bs'}^{\dagger} d_{as}
c_{\beta\sigma'}^{\dagger} c_{\alpha\sigma},
\label{eq:RG1}
\end{equation}
where 
\begin{eqnarray}
\label{eq:RG2}
\mathcal{V}_{bs',as}^{\beta\sigma',\alpha\sigma}\!\!&=&\!\!
\frac{1}{4}\tau_{\sigma\sigma'}^z\tau_{ss'}^z
(J_s^z)_{ba}^{\beta\alpha}+\frac{1}{8}(\tau_{\sigma\sigma'}^+\tau_{ss'}^-+
\tau_{\sigma\sigma'}^-\tau_{ss'}^+)(J_s^\bot)_{ba}^{\beta\alpha}\nonumber\\*
&+&\!\!\frac{1}{4}\delta_{\sigma\sigma'}
\delta_{ss'}\tau_{ab}^x (P_s^x)_{\bar{a}a}^{\beta\alpha}
-\frac{1}{4}\delta_{\sigma\sigma'}
\delta_{ss'}\delta_{ab} (P_s^0)_{\bar{a}\bar{a}}^{\beta\alpha}.
\end{eqnarray}
Initially both interaction terms \eqref{eq:SW29} and \eqref{eq:SW30}
are rotational symmetric in spin space, i.e. $J^z=J^\perp=P^0=P^x$ are
independent of the spin index $s$ on the dot. As we will see
below, the RG flow leads to a breaking of this symmetry.

Furthermore, from now on we use the Einstein summation convention and
parametrise the couplings as $J^{\beta\alpha}_{ba}=r_{\alpha
  b}r_{\beta a}J_0$ with $r_{\alpha b}=t_{\alpha b}/t_{L1}$ and
$t_{L1}=t_{L2}=1$, $t_{R1}=r$, $t_{R2}=1/r$.  We use $\delta$ as unit
of energy ($\delta\equiv 1$), $r$ as asymmetry parameter and set
$\hbar=k_B=e=g\mu_B=1$.

\section{PT and RG equations}

We first study the system using second order PT.  The current is
computed using a rate-equation approach (see e.g. ref.~\cite{S}).
Following ref.~\cite{Paaske} we use a pseudofermion
representation~\cite{Pseudofermion} for the Kondo spins; for details
of the calculation we refer to ref.~\cite{Mueller}. We define the dot
Green's function, $ G_{as}(\tau,\tau') = - i \langle T_C d_{as}(\tau)
d_{as}^\dag(\tau') \rangle$, where $T_C$ is the contour ordering of
the times $\tau, \tau'$ on the Keldysh contour. For the lesser Green's
function we assume $G^<_{as}(\omega) = i n_{as} A_{as}(\omega)$, where
$A_{as}(\omega)$ denotes the non-interacting spectral function given
by a $\delta$-function at $\varepsilon_{as}$.  The stationary
occupation of the state $as$ is determined by the rate equation
\begin{equation}
  \label{eq:rate_eq}
  \sum_{bs'} \Big( \Gamma_{as \to bs'} n_{as} 
  - \Gamma_{bs' \to as} n_{bs'}  \Big) = 0.
\end{equation}
The rates are defined via the retarded self energy as
$\sum_{bs'}\Gamma_{as\to bs'} \equiv \Gamma_{as}(\varepsilon_{as}) = -
2\, \mathrm{Im}[\Sigma^R_{as}(\varepsilon_{as})]$. In second order PT
(Fermi's golden rule) we find
\begin{equation}
\Gamma_{as}(\varepsilon_{as})=
2\pi\left|N_0\mathcal{V}_{as;bs'}^{\beta\sigma';\alpha\sigma}\right|^2 
w(\mu_\alpha-\mu_\beta+\varepsilon_{bs'}-\varepsilon_{as}).
\label{eq:Gamma_PT}
\end{equation}
Here $w(x) = x n_B(x)$, with the finite-temperature Bose function $n_B(x)$, gives the
energy window allowed for transitions. We note that the
self-consistency equations~\eqref{eq:rate_eq} have to be solved under
the constraint $\sum_{as}n_{as} = 1$.  Results of the perturbative
treatment of the occupation numbers and the differential conductance
are shown in figs.~\ref{fig:fig3} and~\ref{fig:fig4}.

In order to include Kondo correlations we use a non-equilibrium poor
man's scaling approach with frequency-dependent coupling
functions~\cite{Rosch}.  The underlying approximations are devised for 
 ${\rm max}\, (V,T,B)\gg T_K$ providing a cutoff for the flow to strong coupling.
We start with the representation
\eqref{eq:RG1}. During the RG flow the vertex $\mathcal{V}$ acquires a
dependence on the energies of the incoming and outgoing lead electrons
as well as the involved dot states. Approximating the latter by their
respective resonance energies $\varepsilon_{as}$ and using energy
conservation we deduce that $\mathcal{V}$ will depend on the frequency
$\omega$ of the incoming electron only. Thus during the RG flow we can
use the parametrisation \eqref{eq:RG2} with $\omega$-dependent
couplings $J^z$, $J^\perp$, $P^0$ and $P^x$.  Explicitly, the RG
equation for the vertex reads
\begin{equation}
\begin{split}
&\frac{\partial}{\partial\ln
  D}\,\mathcal{V}_{bs',as}^{\beta\sigma',\alpha\sigma}(\omega)=\\
&\;\;\Theta_{\omega+\varepsilon_{as}-\varepsilon_{cs''}-\mu_\gamma}
\mathcal{V}_{bs',cs''}^{\beta\sigma',\gamma\sigma''}(\mu_\gamma)
\mathcal{V}_{cs'',as}^{\gamma\sigma'',\alpha\sigma}
(\mu_\gamma\!+\!\varepsilon_{cs''}\!-\!\varepsilon_{as})\\
&-\Theta_{\omega-\varepsilon_{bs'}+\varepsilon_{cs''}-\mu_\gamma}
\mathcal{V}_{bs',cs''}^{\gamma\sigma'',\alpha\sigma}
(\mu_\gamma\!-\!\varepsilon_{cs''}\!+\!\varepsilon_{bs'})
\mathcal{V}_{cs'',as}^{\beta\sigma',\gamma\sigma''}(\mu_\gamma),\label{eq:RG}
\end{split}
\end{equation}
where $\Theta _{\omega }=\Theta (D-|\omega +\mathit{i}\Gamma |)$,
$D<D_0$ is the running cutoff and $\Gamma$ is the decoherence rate
cutting off the RG flow \cite{Rosch}. Motivated by the expressions
entering the computation of the current we use the mean value of the
rates, $\Gamma=\frac{1}{4}\sum_{as}\Gamma_{as}$.  Generalising the
rate \eqref{eq:Gamma_PT} from PT using frequency-dependent coupling
functions, we find
\begin{equation}
\begin{split}
&\Gamma_{as}(\varepsilon_{as})=\frac{1}{2}
\int\limits_{\mu_\alpha}^{\mu_\beta+\varepsilon_{as}-\varepsilon_{bs'}}
\mathrm{d}\omega\,
\frac{\left|N_0\mathcal{V}_{as;bs'}^{\beta\sigma';\alpha\sigma}(\omega)
\right|^2}
{\mu_\beta+\varepsilon_{as}-\varepsilon_{bs'}-\mu_\alpha}\\
&\times\int\mathrm{d}\omega'A_{bs'}(\omega')
\int\mathrm{d}\omega''
\bigl(1-f^\alpha(\omega'')\bigr)
f^\beta(\omega''-\varepsilon_{as}+\omega') \label{eq:Gamma1a}
\end{split}
\end{equation}
with the Fermi function $f^{\alpha }(\omega)=1/(1+e^{(\omega -\mu
  _{\alpha })/T})$. Here we have used that the spectral function
$A_{as}(\omega)$ is given by a Lorentzian of width $\Gamma$ centred
around $\omega= \varepsilon_{as}$ (replacing the $\delta$-function)
and thus is strongly peaked. The convolution of the Fermi functions
> with broadened spectral functions leads to an effective temperature cutting off
> the flow to strong coupling at low bias voltages~\cite{Paaske}. 
The equations \eqref{eq:RG} and
\eqref{eq:Gamma1a} have to be solved self-consistently; we find a good
convergence within a few iterations.

\begin{figure}[t]
\begin{center}
\includegraphics[width=0.9\linewidth,clip=true]{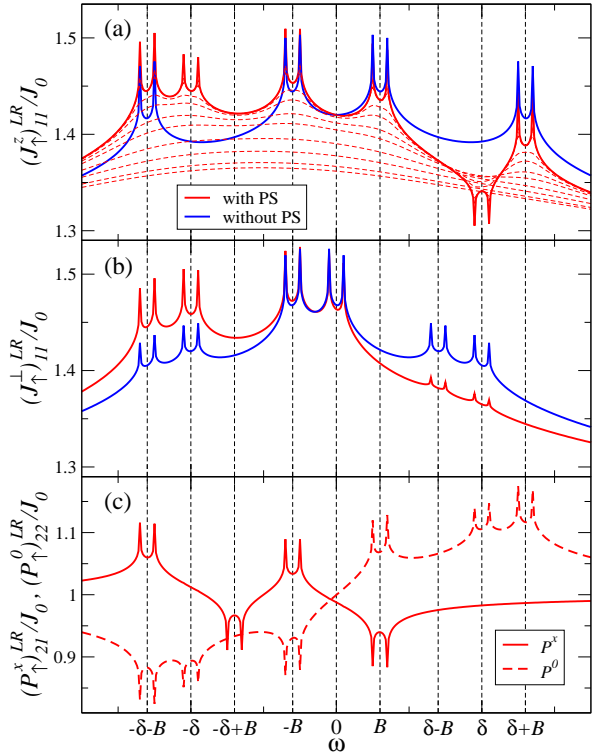}
\caption{(Colour online) Flow of the couplings $J^z$ (a), $J^{\perp}$
  (b) and $P^{0,x}$ (c) as a function of $\omega$ for $r=1$, $B=0.3$,
  $T=0.01$, and $V=0.1$; the initial values are $N_0J_0=0.01$ and
  $D_0=1000$. Results with (red) and without (blue) orbital PS are
  compared. The evolution of $J^z$ under the RG flow is indicated by
  dashed red lines; the final value is obtained at $D=\Gamma\approx
  0.0002$. We note that the RG flow results in $J^z\neq J^\perp$ and that
  the orbital PS yields new structures in $J^z$ at $\omega=-\delta\pm V/2$ and
  $\omega=\delta\pm V/2$. The orbital PS is only weakly renormalised.}
\label{fig:fig2}
\end{center}
\end{figure}

Typical results for the flow of the frequency-dependent Kondo
couplings $J$ are shown in fig.~\ref{fig:fig2}, where we focus on one
representative component of $J^{z,\perp}$ and $P^{0,x}$. In order to
assess the importance of the orbital PS contributions, a comparison
between results with and without orbital PS is provided.  Lowering the
cutoff down to the self-consistently determined rate $\Gamma$ leads to
an overall increase of $J$ with respect to the bare initial couplings
$J_0$. Furthermore, pronounced peaks develop in correspondence of
resonant transport involving spin- and orbital-flip processes. These
peaks are split due to the finite bias by $\pm V/2$.  The singular
behaviour is cut off by $\Gamma$; the induced noise due to the
stationary current leads to a finite life time of the dot
states~\cite{Rosch}.  This behaviour as well as the position of the
resonances can be inferred from the analytic structure of the flow
equations.  In the following we focus on the coupling $J^z$, the RG
equations of the other couplings can be found in ref.~\cite{Mueller}.
Explicitly, the RG equation for one representative component of $J^z$
in the absence of orbital PS reads
\begin{eqnarray}
&&\hspace{-12mm}
\frac{\partial (J_\uparrow^z)_{11}^{L R}(\omega)}{\partial\ln D}=\nonumber\\*
&&\hspace{-10mm}
-\frac{1}{4}\left[(J_\uparrow^\bot)_{11}^{\gamma R}
(\mu_\gamma+B)(J_{\downarrow}^\bot)_{11}^{L\gamma}(\mu_\gamma)
\Theta_{\omega-\mu_\gamma-B}\right.
\nonumber\\*[2mm]
&&\hspace{-7mm}+(J_\uparrow^\bot)_{11}^{L\gamma}(\mu_\gamma)
(J_{\downarrow}^\bot)_{11}^{\gamma R}(\mu_\gamma-B)\Theta_{\omega-\mu_\gamma+B}
\nonumber\\*[2mm]
&&\hspace{-7mm}+(J_\uparrow^\bot)_{21}^{\gamma R}(\mu_\gamma+\delta+B)
(J_{\downarrow}^\bot)_{12}^{L\gamma}(\mu_\gamma)
\Theta_{\omega-\mu_\gamma-\delta-B}
\nonumber\\*[2mm]
&&\hspace{-7mm}\left.+(J_\uparrow^\bot)_{21}^{L\gamma}(\mu_\gamma)
(J_{\downarrow}^\bot)_{12}^{\gamma R}(\mu_\gamma-\delta-B)
\Theta_{\omega-\mu_\gamma+\delta+B}\right].
\label{eq:RG15}
\end{eqnarray}
Characteristic Kondo divergences appear at $\omega=\pm (\delta+B)$ and
$\omega=\pm B$, which are split by $\pm V/2$ in presence of a finite
bias voltage.  Similarly to the conventional Kondo model, $J^z$ is
renormalised by two spin-flip processes $\propto
(J^\perp)^2$. Schematically, the RG equation has the form $\partial
J/(\partial \ln D )= -2J^2$ thus leading to a divergence at low
energies.  

Qualitatively new effects arise in the presence of orbital PS contributions,
which lead to the following additional terms on the right-hand side of
\eqref{eq:RG15} 
\begin{equation}
  \begin{split}
    &-\frac{1}{4}(J_\uparrow^z)_{21}^{L\gamma}(\mu_\gamma)
    (P_\uparrow^x)_{12}^{\gamma R}(\mu_\gamma-\delta)
    \Theta_{\omega-\mu_\gamma+\delta}\\
    &+\frac{1}{4}(J_\uparrow^z)_{21}^{\gamma
      R}(\mu_\gamma+\delta)(P_\uparrow^x)_{12}^{L\gamma}(\mu_\gamma)
    \Theta_{\omega-\mu_\gamma-\delta}+\{J^z\leftrightarrow P^x\}.
  \end{split}
\label{eq:RG16}
\end{equation}
$P^x$ always involves a change of the orbital index with energy $\pm
\delta$.  As can be seen in fig.~\ref{fig:fig2}c the orbital PS
couplings are only weakly renormalised.  For the symmetric case the
orbital PS couplings $P$ do not flow at lowest order, they only evolve
due to the finite frequency-dependence acquired by the Kondo couplings
during the flow. Moreover, the averages of $P^x$ entering the
expressions for the rates \eqref{eq:Gamma1a} and the current
\eqref{eq:current} will only slightly differ from the bare values and
thus induce only minor effects.  However, the flow of $P$ feeds back
into the flow of the Kondo couplings where it results in qualitatively
new features.  As shown in fig.~\ref{fig:fig2}a, for $\omega=\delta\pm
V/2$ the flow of $J^z(\omega)$ exhibits a crossover from a gradual
increase to a \emph{dip} as $\Gamma$ decreases.  In this regime the
flow equation assumes the form $\partial J/(\partial\ln D)=JP/2$; the
reversed sign with respect to the characteristic Kondo divergence
gives rise to the observed dips.  These are not present for $B=0$ as
they merge with the larger peak coming from $\delta+B$. In this case
the orbital PS terms only induce an asymmetry as a function of
$\omega$.  This can be observed for $J^\perp$ (see
fig.~\ref{fig:fig2}b), where the contribution of the orbital PS only
changes the weight of already resonant processes.  $J^\perp$ is
renormalised by $J^\perp J^z$ and $J^\perp P$, where both $J^z$ and
$P$ do not involve spin-flip processes and thus no difference appears
at finite magnetic field.  For asymmetric couplings $r>1$ we find that
existing peaks at $\omega=0,\pm B$ are enhanced while all others are
strongly suppressed.

\section{Current and dot occupation} 

The current is obtained by taking into account the generated frequency
dependence of the couplings. Specifically we find
\begin{eqnarray}
  I\!\!&=&\!\!
  \frac{e}{2\hbar}\;\int\limits_{-\varepsilon_{bs'}+
    \varepsilon_{as}+\mu_R}^{\mu_L}\mathrm{d}\omega\,
  \frac{\left|N_0\mathcal{V}_{as;bs'}^{R\sigma;L\sigma'}(\omega)\right|^2}
{\mu_L+\varepsilon_{bs'}-\varepsilon_{as}-\mu_R}\nonumber\\
  &&\!\!\times\!\int\mathrm{d}\omega'\mathcal{A}(\omega')\!
  \int\mathrm{d}\omega''
  \Bigl[n_{bs'}\bigl(1-f^R(\omega'\!+\!\omega'')\bigr)
  f^L(\omega'')\nonumber\\
  &&\qquad\qquad\qquad\,\,\qquad+n_{as}f^R(\omega'\!+\!\omega'')
  \bigl(1-f^L(\omega'')\bigr)\Bigr]\nonumber\\
  &&\!\!+\{{\rm same}\;\, {\rm expression}\; \,{\rm with}\;\, L\leftrightarrow R\},
\label{eq:current}
\end{eqnarray}
with $\mathcal{A}(\omega)$ a Lorentzian\footnote{$\mathcal{A}(\omega)$
  is given by the convolution of $A_{as}(\omega)$ and
  $A_{bs'}(\omega)$.} of width $2\Gamma$ centred around
$\varepsilon_{bs'}-\varepsilon_{as}$.  Given the complicated form of
the RG equations a detailed analytical understanding of the behaviour
of the physical observables is not feasible, especially as weighted
integrals over the coupling functions are involved.  Thus we have
evaluated the dot occupation numbers, the decoherence rates and the
current numerically. The renormalisation of the energy levels and the magnetic
field are neglected in the present analysis, since they turn out to
contribute only at higher order.

\begin{figure}[t]
\begin{center}
\includegraphics[width=.9\linewidth,clip]{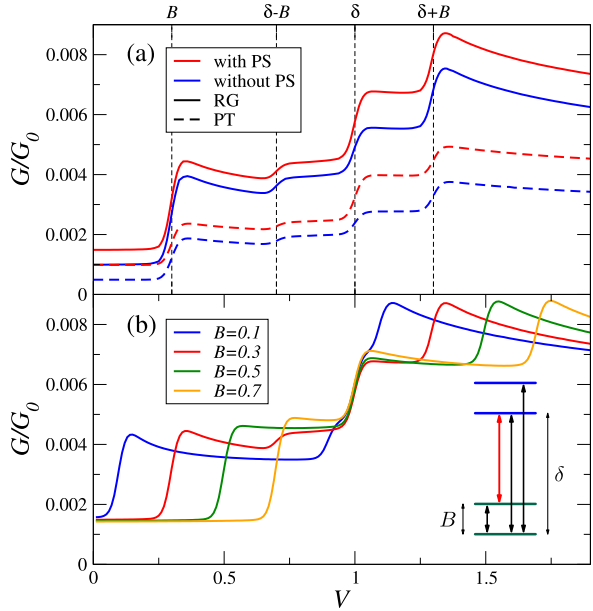}
\caption{(Colour online) Differential conductance,
  $G=\mathrm{d}I/\mathrm{d}V$, normalised to $G_0=2e^2/h$, for symmetric
  couplings ($r=1$); a) RG results (full lines) are compared to PT (dashed
  lines), both with (red) and without (blue) orbital PS for parameters as in
  fig.~\ref{fig:fig2}; b) RG results including orbital PS for the same
  parameters but different values of $B$. The cascade step is visible at
  $V=\delta-B$ provided $B<\delta/2$; Inset: schematic representation of the
  possible transitions between the dot orbitals, the cascade transition being
  marked in red.}
\label{fig:fig3}
\end{center}
\end{figure}
First we present results for the symmetric situation $r=1$ and discuss
the effects of a finite magnetic field. Fig.~\ref{fig:fig3}a shows the
differential conductance obtained from PT and RG calculations both
with and without orbital PS.  As long as the couplings are symmetric,
the differential conductance is symmetric with respect to the voltage
and thus only $V>0$ is shown. We observe that the inclusion of the
orbital PS leads to an enhancement of the conductance, which is simply
due to the increase of transport channels.  This enhancement is
voltage independent except that its value changes at $V=\delta$, which
indicates the importance of the coupling of the orbital PS terms to the $J^z$
processes.  Regarding the effect of the RG we observe an overall
increase of the conductance, which is due to the increase of the Kondo
couplings under the RG flow (see fig.~\ref{fig:fig2}). Furthermore,
there is a clear enhancement of the inelastic cotunneling signatures
at $V=B$ and $\delta+B$ involving spin-flip processes. In contrast,
the effect of the orbital PS remains unchanged under the RG flow.

\begin{figure}[t]
\centering
\includegraphics[width=0.9\linewidth,clip]{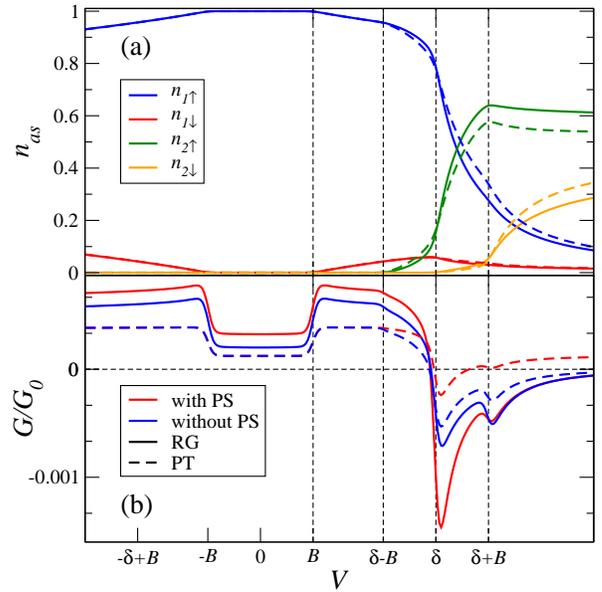}
\caption{(Colour online) Dot occupation numbers (a) and differential
  conductance (b) for asymmetric couplings with $r=5$, $N_0J_0=0.005$
  and other parameters as in fig.~\ref{fig:fig2}.  For the occupation
  numbers RG results (full lines) are compared to PT (dashed lines)
  both including orbital PS.}
\label{fig:fig4}
\end{figure}

The cascade step~\cite{Schmaus} at $V=\delta-B$ has its physical
origin in the non-equilibrium dot occupations (see
fig.~\ref{fig:fig4}a). For $V>B$ the orbital $1\!\!\!\downarrow$ gets
populated from which the next higher orbital $2\!\!\uparrow$ is split by
$\delta-B$ (see inset of fig.~\ref{fig:fig3}b). Therefore, a finite
population of $2\!\!\uparrow$ already sets in at $V=\delta-B$, which
results in the cascade step in the conductance displayed in
fig.~\ref{fig:fig3}.  This is in analogy to the Balmer series
observed in the traditional spectroscopy of hydrogen, where instead of
transitions from the ground state (Lyman series) the transitions from
an excited state are probed.  As one can observe in
fig.~\ref{fig:fig3}a the cascade resonance is only weakly enhanced by
RG. The reason might be that the RG only starts to play a role at
higher orders since the cascade effect is due to transitions from an
excited state.  Clearly the cascade effect is not present at $B=0$
while for $B=\delta/2$ it is superimposed by the inelastic cotunneling
process at $V=B$ (see fig.~\ref{fig:fig3}b).

In the following we discuss the effect of asymmetric couplings and the
resulting NDC.  Fig.~\ref{fig:fig4} shows PT and RG results for the
dot occupations and differential conductance for asymmetric couplings
with $r=5$. This implies that $t_{R1}\gg t_{R2}$, i.e. the transport
through the second orbital is strongly suppressed. At $V\approx\delta$
we observe a strong increase of $n_{2\uparrow}$ at the cost of
$n_{1\uparrow}$ which consequently leads to a decrease in the current
and thus the appearance of NDC. The increase of $n_{2\downarrow}$
similarly yields a side peak located at $V\approx\delta+B$.  A simple
explanation for the appearance of the NDC is provided by PT in the
limit of $r\gg 1$, $B=0$ and $T=0$. Introducing the orbital
polarisation $M_{12}=\frac{1}{2}
\sum_{\sigma}(n_{1\sigma}-n_{2\sigma})$ we obtain for the differential
conductance at $V\approx\delta$
\begin{equation}
  \frac{G}{G_0}\propto 2+3M_{12}+
  \delta\frac{\mathrm{d}M_{12}}{\mathrm{d}V}\bigg|_{V=\delta}.
\label{eq:currpt}
\end{equation}
This shows that the NDC originates in the voltage-dependent occupation
numbers and that a sufficiently strong occupation inversion
($M_{12}<0$) and/or sufficiently fast reduction of the orbital
polarisation ($\mathrm{d}M_{12}/\mathrm{d}V<0$) are necessary.  We
note that the second orbital is not occupied for negative voltages,
thus NDC is absent in this regime.  For generic $B$ and $r$ NDC shows
up in the vicinity of $n_{1\uparrow}=n_{2\uparrow}$ in the interval
$\delta<V<\delta+B$.  NDC is favoured for small magnetic fields, as
the occupation inversion becomes less feasible with increasing Zeeman
splitting.  While the dot occupations do not exhibit remarkable
renormalisation effects with respect to the PT results (see
fig.~\ref{fig:fig4}a), the conductance is strongly affected by both
renormalisation effects and orbital PS contributions.  The NDC is
strongly enhanced by the RG, and in contrast to the case of symmetric
couplings (see fig.~\ref{fig:fig3}a) the orbital PS has opposite
effects on the PT and RG results: while the orbital PS reduces the NDC
in PT, it induces an increase in the RG results.  Depending on the
initial system parameters the NDC might even exclusively be generated
during the RG flow.  At low bias voltage the effect of the orbital PS
contributions is strongly reduced in PT, from the analytic
calculations this effect can be shown to be proportional to $(P^0)^2
\sim 1/r^2$.  In contrast, the corresponding RG results show a marked
$r$- and $V$-dependence of the orbital PS contributions.  Concerning
the cascade effect the onset of $n_{2\uparrow}$ becomes more
pronounced with stronger asymmetry.

We stress that the mechanism leading to NDC in the present model is
fundamentally different from the NDC observed in the interacting
resonant level model~\cite{IRLM} where it has its origin in a
voltage-dependent renormalisation of the hopping amplitudes towards
smaller values.

Finally we note that the temperature considered here is
$T=0.01\,\delta\gg T_K$ with the rough estimate $T_K\sim
D_0\,\exp[-1/(N_0J_0)]$.  For higher temperatures $T\sim 0.1\,\delta$
thermal smearing inhibits a clear distinction of the different
transitions.  The small absolute values of the conductance are due to
the small values of the initial couplings $N_0J_0\sim 0.01$ required
by our perturbative treatment. However, we expect the cascade effect
as well as the appearance of NDC to be generic features of
non-equilibrium transport through multi-orbital quantum dots. In
particular, these features are expected to show up 
in the kind of transport experiments 
on molecules and nanowires as carried out in
Refs.~\cite{ndcexp,Osorio, Roch, Sapmaz, jens,Jespersen} 
where the
couplings to the leads will generically be strongly asymmetric and
orbital dependent.

\section{Conclusion}

We studied a minimal model for a multi-orbital quantum dot exhibiting
the cascade effect and NDC, namely a two-orbital Anderson model with
asymmetric and orbital-dependent hopping amplitudes in a magnetic
field. We considered the regime of strong Coulomb interactions where a
SW transformation leads to an effective Kondo model including orbital
PS terms. The non-equilibrium transport through this system was
studied using a perturbative RG treatment with frequency-dependent
couplings. In particular, the RG flow leads to a strong enhancement of
the NDC. The observed strong influence of the orbital PS on the
scaling of the Kondo couplings indicates an interaction-driven
interplay of spin and orbital degrees of freedom even in absence of
spin-orbit interactions or spin-dependent tunneling amplitudes.
The inclusion of additional spin-orbit interactions represents
an interesting extension beyond the scope of the present analysis.

We are grateful to K.~Flensberg, S.~Jakobs, T.~S.~Jespersen,
J. ~Nyg{\aa}rd, J.~Paaske, and H.~Schoeller for helpful discussions.
We acknowledge kind hospitality at the Niels Bohr Institute,
University of Copenhagen (S.A. and S.M.) and the
Department of Physics, University of Basel (S.M.). This
work was supported by the DFG-FG 723 and 912, and the Robert Bosch
Foundation.


\end{document}